\documentclass[a4paper,10pt]{article}
\usepackage{enumerate}
\usepackage{amsmath}
\usepackage{amsfonts}
\usepackage{amssymb}
\usepackage{amsthm}
\usepackage{graphicx}
\usepackage{geometry}
\usepackage{float}
\usepackage{graphics}
\usepackage{float}
\usepackage{times}
\usepackage{color}
\usepackage{tikz}
\usepackage{fontenc}
\usepackage[english]{babel}
\usepackage{fontenc}
\usepackage[english]{babel}
\theoremstyle{definition}
\newtheorem{Theorem}{Theorem}
\newtheorem{definition}{Definition}

\newtheorem{lemma}{Lemma}
\newtheorem{remark}{Remark}
\newtheorem{example}{Example}

\newtheorem{proposition}{Proposition}

\begin{document}
	\title{Weighted Myerson value for Network games}

	\author{Niharika Kakoty$^{1}$, Surajit Borkotokey$^{1*}$, Rajnish Kumar$^{2}$, Abhijit Bora$^{1}$ }
	\date{} 
	\maketitle
	
	\begin{center}
		$^1$Department of Mathematics, Dibrugarh University, Dibrugarh, 786004, Assam, India\\
		$^2$Queen's Management School, Queen's University, Belfast BT7 1NN, Northern Ireland, UK\\
		$^*$Corresponding authors' e-mail:sborkotokey@dibru.ac.in
	\end{center}
	
  \vspace{1cm}
	\begin{abstract}
	 We study the weighted Myerson value for Network games extending a similar concept for communication situations. Network games, unlike communication situations, treat direct and indirect links among players differently and distinguish their effects in both worth generation and allocation process. The weighted Myerson value is an allocation rule for Network games that generalizes the Myerson value of Network games. Here, the players are assumed to have some weights measuring their capacity to form links with other players. Two characterization of the weighted Myerson value are provided. Finally we propose a bidding mechanism to show that the weighted Myerson value is a subgame perfect Nash equilibrium under a non-cooperative framework.
	\end{abstract}
	
	\noindent\textbf{Keywords}: Myerson value, network games, position value, weighted value.
	
	\noindent\textbf{2010 Mathematics Subject Classification}: 91A12.
	
	\section{Introduction}
	In this paper, we propose the weighted Myerson value for Network games that combines the notion of the weighted Shapley value of cooperative games with transferable utilities (TU games in short) due to \cite{Kalai, Shapley} and the Myerson value of Network games due to \cite{Jackson1996, Myerson}. The Myerson value for Network games is itself a manifestation of the Shapley value of TU games with the assumption that players in the network are symmetric in terms of their capacity to form and sustain links in the network. However, in practice, this assumption is too restrictive, see for example \cite{Haeringer1999, Haeringer}. Thus, we assume that each player in the network is endowed with a weight and accordingly the weighted Myerson value will take care of these weights in allocating the worth generated by the network.
\par Mathematically speaking, a cooperative game with transferable utilities is a pair $(N, v)$ where $N$ denotes the player set and $v$ denotes a characteristic function  $v : 2^N \mapsto \mathbb R$ such that $v(\emptyset)=0$. The domain $2^N$ of $v$ is the power set of $N$ containing all the coalitions of $N$. Thus, $v$ assigns a real number to each coalition which we call its worth. The worth $v(N)$ is the worth of the grand coalition $N$. It is assumed that the grand coalition $N$ is formed in a TU game. The problem is then to allocate the worth of the grand coalition $v(N)$ among the players. Among the many solutions (allocations) proposed from time to time, the Shapley value \cite{Shapley53} is perhaps the most popular value. It is the average of the marginal contributions of the players over all possible orderings of joining a coalition.  The underlying assumption here is that all the coalitions must be formed before the grand coalition is formed at the final stage. However, Myerson \cite{Myerson} in his seminal paper, argues that  coalitions are feasible among the players if they have some kind of pre-existing connections among themselves. This connections can be represented by the edges or links in a graph. Therefore, Myerson terms this new TU game as a graph restricted game. He obtains an allocation rule for the graph restricted game which is the Shapley value of an associate TU game. This value is later called the Myerson value in \cite{Jackson1996}. In \cite{Jackson1996}, the graph restricted game is renamed as communication situation. In a communication situation, it only matters whether players are connected through a graph or not, take for example, the worth of a coalition  does not depend on whether players in it are connected by  direct or indirect links. However, in \cite{Jackson1996}, it is argued that when network formation is costly, take for example, when the physical structure of a network exists, the direct and indirect links have their own significances, and therefore, the worths of the coalitions need to be replaced by the worths of the subnetworks. Thus a Network game is a pair $(N, v)$ where $N$ is the player set but $v$ is a function that assigns a real number to each subnetwork of a given network formed by the players in $N$. The Myerson value of communication situations has a direct counterpart in Network games as well. Other allocation rules for communication situations and Network games such as the position value \cite{Borm, Meesen, Nouweland},  the multilateral value \cite{borejor}, the outside option values \cite{belau3} etc., are also found in the literature, however, in this current study, our focus is on the Myerson value only.
\par In TU games, the weighted Shapley value \cite{Shapley} is proposed as a further generalization of the Shapley value such that while allocating the worth of the grand coalition each player receives her payoff proportional to her pre-assigned weight. The Shapley value, in particular, is one such allocation where the players are considered to be of equal weights. Bringing this idea to communication situations, Haeringer \cite{Haeringer1999, Haeringer} proposes the weighted Myerson value where he assumes that players in a communication situation have, in general, non-identical bargaining power and therefore, the ordinary Myerson value cannot justify a fare allocation for those players. In a similar manner, in Network games, it is quite intuitive to assume that players in the network are un-equal in terms of forming a link or sustain it. To make the case more complicated, we may further assume that the players have un-equal capabilities for forming direct and indirect links and sustain them. Thus a weight system measures this un-equal capacities of players. Accordingly, the weighted Myerson value awards the players their share from the value generated by the given network proportional to their weights. To understand how the weights of the players in a network game play a significant role in allocating the worth of a network we present the following two examples.
\begin{example}
The Belt and Road Initiative (BRI) of China aims to connect Asia, Europe, and Africa through a vast network of land and sea routes. Its main goal is to boost regional cooperation, trade, and economic growth. The BRI is attractive because it involves significant investments in various infrastructure projects like ports, roads, railways, airports, power, and telecommunication networks. With around 156 countries taking part in the BRI, each with different economic strengths and negotiation abilities, the way they participate varies.  In this context, a country's weight represents its influence in negotiations. Typically, larger and more advanced countries have more negotiating power and can secure better trade deals. Smaller and less developed countries may face stricter trade conditions to protect their own industries. The weighted Myerson value ensures a fair distribution of benefits and drawbacks in trade agreements, considering each country's economic strength and negotiation ability. This approach aims for fairness and long-lasting trade agreements, especially when there are economic disparities among countries.
\end{example}
\begin{example}
 Next, take the example of the Paris Agreement, which aims to combat climate change in order to limit global warming. With about 195 countries in this network with different historical and current greenhouse gas emissions, the weights of participating countries represent the level of their greenhouse gas emissions. The countries having higher rates of emissions receive higher weights, reflecting their greater contribution to the climate crisis. When resources and commitments are distributed for climate change, the weighted Myerson value will guarantee that countries with higher emissions (so greater responsibility for this problem) will pay a greater share of their resources. This promotes fairness with respect to the contribution of each country to emissions. When defining global climate policy, countries with less emissions can have a greater say. This encourages countries with greater responsibility for emissions to take proactive measures to reduce their impact on climate change. The weighted Myerson value will ensure fair and coordinated efforts to address climate change within the framework of the Paris Agreement.
\end{example}
\par After the weighted Myerson value is being introduced for Network games, we  characterize it following similar characterizations by \cite{belau3, Haeringer1999}.  We also provide a non-cooperative approach to obtain the value as an SPNE of a suitably designed bidding mechanism in line with \cite{Hart1996,castrillo}	
\par The rest of the paper is organized as follows. In Section 2 we present the definitions and notations that are relevant to the development of the current paper. Section 3 includes two distinct characterizations for the class of weighted Myerson values. In Section 4, we discuss the concept of network potential as introduced by Slikker \cite{Slikker05b} and present a characterization of the weighted Myerson value using this notion. Section 5 contains two seperate bidding mechanisms designed for the class of weighted Myerson values, highlighting their practical implications and applications. In Section 6, we outline potential areas for future research and exploration within this domain.	
\section{Preliminaries}\label{Section2}
In this section, we compile the notations, definitions and results to be used in our current paper from \cite{belau3, Haeringer1999, Jackson2005}.
	   \begin{itemize}
			\item \textit{Players}
			\newline Let $N= \{1, 2, ..., n\}$ denote the set of players. A coalition $S$ is a subset of $N$. We use the corresponding lower case letter $s$ to denote the cardinality of coalition $S$, thus $|S| = s$.
			
			\item \textit{Networks}
			\newline In a set of players denoted by $N$ where each player is distinct, a link represented as $ij$ signifies an undirected connection between player $i$ and player $j$. It is important to note that $ij$ and $ji$ are equivalent and represent the same relationship.
			\par Let  $g_{N}=\{ij$ $|$ $i,j\in N$, and $i\neq j\}$ be the set of all possible links with the player set $N$. This set is known as the complete network. The set of all possible networks on $N$ is given by $\mathbb G^N=\{ g \mid g\subseteq g_{N} \}$. The network $g_0=\emptyset$ is the network without any links, which we call the empty network. Let $|g|$ denote the number of links in a network $g$. It follows that, $|g_N|= \binom{n}{2} = \tfrac{1}{2} n(n-1)$, and $|g_0|=0$. 
			
			\par The set of players in the network $g$ is denoted by $N(g)$. For every network $g\in\mathbb G^N$, and every player $i\in N$ we denote the neighborhood of $i$ in $g$ by $N_{i}(g)=\{ij\in g:i\neq j\}$. Thus, $N_i(g)$ is the set of players with whom $i$ is directly linked in $g$.  Also, denote by $n_{i}(g)=|N_{i}(g)|$, and $n(g)=|N(g)|$. Denote by $g_i=\{ij\in g\mid j\in N_{i}(g)\}\subseteq g$ the set of links of player $i$ in $g$. \\ 
			
			\item \textit{Networks on subsets of players}
			\newline For any $g\in\mathbb G^N$, and $S\subseteq N$, the restriction of $g$ on the coalition $S$ is denoted by $g|_{S}$, and is given by $g|_{S}=\{ ij \in g \mid i,j \in S \}$. 
			
			\item \textit{Paths}
			\newline  A \textit{path} in $g$ connecting $i$ and $j$ is a set of distinct players $\{i_{1},i_{2},\ldots,i_{p}\}\subseteq N(g)$ with $p\geqslant2$ such that $i_{1}=i$, $i_{p}=j$, and $\{i_{1}i_{2},i_{2}i_{3},\ldots,i_{p-1} i_{p}\}\subseteq g$. We say $i$ and $j$ are connected to each other if a path exists between them and they are disconnected otherwise.
		
			\item \textit{Components}
			\newline The network {$h\subseteq g$} is a \textit{component} of $g$ if for all { $i,j\in N(h)$}, $i\neq j$, there exists a path in  {$h$} connecting $i$ and $j$ and for any {$i\in N(h) $} and $j\in N(g)$, $ij\in g$ implies {$ij\in h$}. In other words, a component  of $g$ is simply a maximally connected sub-network of $g$. We denote the set of components of the network $g$ by $C(g)$.
			
			\item \textit{Isolated players} 
			\newline The players in $N$ which are not connected to any players in $g$ are called \textit{isolated players}. Let $N_0(g)$ denote the set of all such isolated players. Then, we have $N_{0}(g)=N\setminus N(g)=\{i\in N\mid N_{i}(g)=\emptyset\}$. Clearly, $N_0 (g_0) = N$.
			
			\item \textit{Value functions}\\
			A \textit{value function} on $\mathbb G^N$ is a function $v \colon\mathbb G^N \to \mathbb R$ such that $v(g_0)=0$. The set of all possible value functions with the player set $N$ denoted by $\mathbb V^N$ is a {$2^{\dfrac{n(n-1)}{2}}-1$} dimensional vector space. Let $v_{0}$ denote the null value function given by $v_{0}(g)=0$ for all $g \in \mathbb G^N.$ 
			\item \textit{A Basis for value functions}
			\newline An important subclass of the class of value functions in $\mathbb V^N$ is the class of unanimity value functions which is defined as follows.
			\begin{itemize}
				\item For any network $g \in \mathbb G^N$ and $\bar g \subseteq g$, the unanimity value function $u_{\bar g}$ is defined $\forall g' \subseteq g$ as: 
				{\[ u_{\overline{g}}(g\prime)=\begin{cases} 
						1 & ~ \text{if} ~ \overline{g} \subseteq g' \\
						0& ~\text{otherwise}.  
					\end{cases}
					\]}
			\end{itemize}
			This class of value functions forms a basis for $\mathbb V^N$. So, each $v\in \mathbb V^N$ such that $v \ne v_0$, can be represented as
						\begin{equation}\label{eq1}
							v=\sum_{g' \in \mathbb G^N}\lambda_{g'}(v)u_{g'}
						\end{equation} 
						where $\lambda_{g'}(v)$ are unique unanimity coefficients corresponding to the unanimity value functions $u_{g'}$ called Harsanyi dividends\footnote{See Page 523 of \cite{Caulier}.} and are given by, $$\lambda_{g'}=\sum_{g'\subseteq g}(-1)^{|g|-|g'|}v(g').$$ 
			\item \textit{Component additive value functions}\\
			An interesting subclass of value functions is that of \textit{Component additive value functions} where the values do not spill over from one component to the other. Thus, a value function $v\in \mathbb V^N$ is component additive if 
			$$v(g) = {\sum_{h \in C(g)} v(h)}\;\;\;\textrm{for any}\;\;g\in \mathbb G^N.$$
			The space of the component additive value functions with player set $N$ is denoted by $\mathbb H^N$, which is a  $\sum_{h \in C(g)}(2^{|h|}-1)$ dimensional subspace of  $\mathbb V^N$. It is shown in \cite{Nouweland} that the component
			additive unanimity value functions forms a basis of the space of
			component additive value functions. So, each $v\in \mathbb H^N$ such that $v \ne v_0$, can be represented as
			\begin{equation*}
				v=\sum_{g' \in \mathbb G^N}\beta_{g'}(v)u_{g'}
			\end{equation*} 
			where $\beta_{g'}(v)$ are unique unanimity coefficients corresponding to the component additive unanimity value functions $u_{g'}$\footnote{for more details, see~\cite{Nouweland}.}. 
\newline An important result concerning component additive value functions is given  in \cite{Nouweland}.
\begin{lemma}[(\cite{Nouweland}, page 267, Lemma 1)]\label{lemma1}\textit{For a component additive value function $v$, $\beta_{g'}(v)=0$ for any network $g\in \mathbb{G}^N$ that has at least two links that belong to two different components.}
\end{lemma}			
		\item \textit{Network game}\\
		A \textit{Network game} is a pair $(N, v)$ where $N$ is the player set and $v$ is a value function.
		    \item \textit{Allocation rules}
			\newline A \textit{network allocation rule} is a function $Y\colon\mathbb G^N \times \mathbb V^N \to \mathbb R^N$ such that for every $g \in \mathbb G^N$ and every $v \in \mathbb V^N$, $Y_i(g,v)= 0$ whenever $i \in N_0(g)$.
		
			\item \textit{Component Balanced allocation rules}
			\newline An allocation rule $Y$ is \textit{Component Balanced} if for every $v \in \mathbb H^N$ and $g \in \mathbb G^N$ it holds that $$\sum_{i\in N(h)}Y_{i}(g,v)=v(h), \;\textrm{for every component $h \in C(g)$.}$$
			Component Balance ensures that there are no externalities across components due to a component additive value functions. So, the value of each component is allocated among the members of the component.
			\item \textit{Equal Bargaining Power}
			\newline An allocation rule $Y$ satisfies \textit{Equal Bargaining Power} if for every network $g \in \mathbb G^N$,  $v \in \mathbb H^N$, and for all  $i,j \in N(g)$, we have
			\begin{equation*}
				Y_{i}(g,v)-Y_{i}(g-ij,v)=Y_{j}(g,v)-Y_{j}(g-ij,v).
			\end{equation*}
Note that achieving {Equal Bargaining Power} doesn't strictly imply that players need to evenly distribute the marginal value of a link. Instead, it signifies that two players should both reap equal benefits or bear equal consequences when the link joining these two players is removed from the network.
\item \textit{Additivity}
\newline For each pair of value functions $v, v' \in \mathbb V^N$ and $g\in \mathbb G^N$, if 
$$Y_{i}(g, v + v')=Y_{i}(g, v)+ Y_i(g, v')~~\forall i\in N,$$ then $Y$ is said to satisfy \textit{Additivity}.
\newline It states that the allocation rule exhibits no externalities when players within a network engage in multiple games, each described by their respective value functions.
\item \textit{Superfluous Link}
\newline A link $l \subseteq g$ is \textit{Superfluous} in $g$ with respect to the value function $v \in \mathbb V^N$, namely,
$$v(g')=v(g'\setminus l)$$ for all $g'\subseteq g$ such that $l \subseteq g'$. 
\newline A \textit{Superfluous Link} exerts no influence on any network.
\item\textit{Superfluous Link Property}
\newline An allocation rule $Y$ on $\mathbb G^{N}\times \mathbb V^{N}$ satisfies the \textit{Superfluous Link Property} if
$$Y(g,v)=Y(g\setminus l,v)$$ for all $g \in \mathbb G^N$ and $v \in \mathbb V^N$ and all links $l$ that are \textit{Superfluous} in $g$ with respect to $v$.
\newline This axiom asserts that the existence or nonexistence of a link has no impact on the value of any network, nor does it influence the allocations of players within the network.
\item \textit{Point Anonymity}
\newline The value function $v$ is \textit{Point Anonymous} if $v(g_{|_{S}})=v(g_{|_{S'}})$ for all $S,S'\subseteq N(g)$ such that $|S|=|S'|.$
\newline Under a \textit{Point Anonymous} value function all active players are equally significant and only the size of the sub-coalitions matter to generate the value of the network and its sub-networks.
\item \textit{Point Anonymity Property}
\newline An allocation rule satisfies \textit{Point Anonymity Property} if for all \textit{Point Anonymous} value functions $v$ and $g \in \mathbb G^N$, there exists $\alpha\in \mathbb{R}$ such that 
	{\[ Y_i(g,v)=\begin{cases} 
						\alpha & ~ \text{if} ~ i \in N(g)\\
						0& ~ \text{otherwise}.  
					\end{cases}
					\]} 
For a \textit{Point Anonymous} value function, the players are not distinguishable and \textit{Point Anonymity} ensures same share for each of the players in the network.
\item \textit{The Myerson value}
\par The Myerson value given in \cite{Jackson1996} is the network allocation rule $Y^{MV} \colon \mathbb G^N \times \mathbb V^N \to \mathbb R^N$ given by
			\begin{equation}\label{eq:myerson}
				Y_{i}^{MV}(g,v)={\sum_{S\subseteq N\setminus i}} \left(v(g|_{S\cup i})-v(g|_{S})\right)  \left( \frac{s!(n(g)-s -1)!}{n(g)!}\right).
			\end{equation}
			for every $g \in \mathbb G^N$ and $v \in \mathbb V^N$.
	\par Another form of the Myerson value using Harsanyi dividends is as follows, see \cite{slikker07}:
	\begin{equation}\label{eq:harsanyi}
					Y_{i}^{MV}(g,v)={\sum_{g' \subseteq g: i \in N(g')}} \dfrac{\lambda_{g'}(v)}{n(g')}.
				\end{equation}
				where $\lambda_{g'}(v)=\sum_{g' \subseteq g}(-1)^{|g|-|g'|}v(g')$ and $n(g')$ is the number of players involved in network $g'.$
			
	\par The Myerson value is the network allocation rule that aggregates the marginal contributions of a players prescribed by the value function. Among many characterizations of the Myerson value, the following characterizations due to \cite{belau3, Jackson2005} are important for us for the development of the weighted Myerson value.
	\end{itemize}
	\begin{Theorem}[\cite{Jackson2005}, page 134, Theorem 1]\label{them:Myerson1}
	\textit{The Myerson value is the unique allocation rule on $G^ N\times \mathbb H^N$ that is Component Balanced and satisfies Equal Bargaining Power.}
	\end{Theorem}
	
	\begin{Theorem}[\cite{belau3}, page 80, Theorem 2]\label{them:Myerson2}
	\textit{The Myerson value for Network games
		on the domain $G^N\times \mathbb H^N$ is an unique allocation rule that is Component Balanced and satisfies Additivity, the Superfluous Link Property and Point Anonymity Property.}
	\end{Theorem}
\section{The Weighted Myerson Value for Network games}
	In this section, we introduce the weighted Myerson value for Network games. First, we assign a positive real number to each player in the network, which forms an- $|N|$ vector of weights. As already mentioned in the introduction, these weights reflect the player's significance in forming and sustaining a link in the network. This can be for example, a measure that determines how much a player is central to the network. Thus, in particular, different centrality measures in networks can be weights in a Network game. For details on various centrality measures in networks, we recommend \cite{bloch,das,freeman}.
	\newline Formally speaking, the weight function is an injective function from the player set to the set of positive real numbers, i.e., $w:N\rightarrow \mathbb R^{+}$ such that $w_i$ is the weight associated with player $i$. Denote by $\mathbb{W}^{N}$ the class of all possible weight distributions (vectors) over the set of players $N.$ With an abuse of notations, we take the  weight function $w$ as a weight vector $(w_i)_{i=1}^{|N|}$ of $\mathbb{W}^{N}$.

\par Given a weight vector, the weighted Myerson value is an allocation rule that generalizes the Myerson value by allowing the players to split their marginal share in each coalition in proportion to their respective weights.
Formally, the weighted Myerson value is denoted by $Y_i^{w-MV}\colon \mathbb{G}^{N}\times \mathbb{V}^{N}\to \mathbb{R}^{N}$ and defined for every $g \in \mathbb{G}^{N}, v \in \mathbb{V}^{N}$ and $w \in \mathbb{W}^{N}$ by 
	\begin{align}\label{Eq.5}
			Y_i^{w-MV}(g,v)&= \sum\limits_{ g^{'}\subseteq g: i\in N(g')}\frac{w_i}{\sum\limits_{j\in N(g^{'})} w_j} \lambda_{g^{'}}(v)
		\end{align}
	where $\lambda_{g^{'}}(v)$ are the unanimity coefficients defined in Section \ref{Section2}.
  \par It is to be noted that the Myerson value shares the values given by the unanimity coefficients equally among the members of the subnetworks of the given network. On the other hand, the weighted Myerson value shares the values obtained from the unanimity coefficients proportional to the weights of the players. If we consider $w_i=$ constant for all $i \in N,$ then the weighted Myerson value coincides with the Myerson value. In the next section, we provide two axiomatic characterizations of the weighted Myerson value.	
	\subsection{Characterization}
	Now, we characterize the weighted Myerson value using axioms some of which are standard in cooperative game theory and are trivial extensions to their network counterparts, and some are specific to Network games. 
	
	\par We begin with the following lemma.
\begin{lemma}\label{Lemma1}
	The weighted Myerson value satisfies \textit{Component Balance}, i.e., for all $g \in \mathbb G^N$, $w\in \mathbb W^N$, $v \in \mathbb H^N$, and each $h \in C(g)$, we have
	\begin{equation*}
		\sum_{i\in N(h)} Y^{w-MV}_i(g,v) = v(h).
	\end{equation*}
\end{lemma}
	\begin{proof}
		Let $v\in \mathbb{H^{N}}$, $g\in \mathbb{G^{N}}$, $h \in C(g)$ and $w \in \mathbb W^N$. Then, we have
		\begin{align*}
		\sum_{i\in N(h)} Y_i^{w-MV}(g,v) &=\sum_{i\in N(h)}\sum_{g'\subseteq g:i\in N(g')} \dfrac{w_i}{\sum\limits_{j\in N(g^{'})} w_j}\lambda_{g'}(v).
		\end{align*}
	From Lemma \ref{lemma1}, we have $\lambda_{g'}(v)=0$ for any $g'\subseteq g$ that has links that belong to different components. So, we have, 
	\begin{align*}
		\sum_{i\in N(h)} Y_i^{w-MV}(g,v)	&=\sum_{g'\subseteq h:i\in N(g')} \dfrac{\sum_{i\in N}w_i}{\sum\limits_{j\in N(g^{'})} w_j}\lambda_{g'}(v)\\
		    &= \sum_{ g' \subseteq h}{\lambda_{g'}(v)}\\
			&=v(h).
	\end{align*}
	This completes the proof.
	\end{proof}	
\noindent The next axiom is the \textit{Weighted Bargaining Power}. This axiom suggests that the loss or gain by two players after removing their link in the network should be proportional to their respective weights. The axiom goes as follows.
\begin{itemize}
\item \textit{Weighted Bargaining Power:}\label{axiom1}
An allocation rule $Y$ satisfies \textit{Weighted Bargaining Power} if for every network $g \in \mathbb G^N$,  $v \in \mathbb H^N$, $w\in \mathbb W^N$ and for all  $i,j \in N(g)$, we have
			\begin{equation*}
				w_j\big[Y_{i}(g,v)-Y_{i}(g-ij,v)\big]=w_i\big[Y_{j}(g,v)-Y_{j}(g-ij,v)\big].
			\end{equation*}
\begin{lemma}\label{Lemma2}
		The weighted Myerson value satisfies {Weighted Bargaining Power}.
\end{lemma}
	\begin{proof}
		\begin{align*}
			w_j \Big[ Y_i^{w-MV}(g,v)- Y_i^{w-MV}(g-ij,v) \Big] &= w_j\Big[\sum_{g'\subseteq g: i\in N(g')} \frac{w_i}{\sum\limits_{j\in N(g^{'})} w_j}\lambda_{g'}(v) \\
			&~~~~~~~~~~~~~~~~-\sum_{g'\subseteq g-ij: i\in N(g')} \frac{w_i}{\sum\limits_{j\in N(g^{'})} w_j}\lambda_{g'}(v)\Big]\\ &=w_j\Big[\sum_{\substack{g'\subseteq g: i\in N(g')\\j\in N(g')}} \frac{w_i}{\sum\limits_{j\in N(g^{'})} w_j}\lambda_{g'}(v)\Big]
			\end{align*}			
Therefore, we have,			
			\begin{align*}
	w_j \Big[ Y_i^{w-MV}(g,v)- Y_i^{w-MV}(g-ij,v) \Big] 		&=w_i\Big[\sum_{\substack{g'\subseteq g: j\in N(g')\\i\in N(g')}} \frac{w_j}{\sum\limits_{j\in N(g^{'})} w_j}\lambda_{g'}(v)\Big]\\
			&=w_i \Big[Y_j^{w-MV}(g,v) - Y_j^{w-MV}(g-ij,v) \Big].
		\end{align*}
	\end{proof}
Our first characterization theorem of the weighted Myerson value goes as follows.
\begin{Theorem}\label{Theorem3}
		The weighted Myerson value is the unique {Component Balanced} allocation rule on $\mathbb{G}^N\times \mathbb{V}^N$ that satisfies {Weighted Bargaining Power.}
	\begin{proof}
	It has already been shown in Lemma \ref{Lemma1}~and~\ref{Lemma2} that the weighted Myerson value satisfies {Component Balance} and {Weighted Bargaining Power.}
	\newline We show that the weighted Myerson value is the unique allocation rule that satisfies {Component Balance} and {Weighted Bargaining Power.} If possible, let $Y$ and $Y^{w-MV}$ both satisfy these two axioms. Let, $g$ be a network with a minimal number of links such that $Y \neq Y^{w-MV}.$ Then, for $h\in C(g)$ and any pair $i,j\in N(h)$,  we get
		$$Y(g-ij,v)=Y^{w-MV}(g-ij,v). $$
	Then using {Weighted Bargaining Power}, we get	
	\begin{align}\label{eq5}
		Y_i(g,v)- \frac{w_i}{w_j}Y_j(g,v) &=Y_i(g-ij,v)- \frac{w_i}{w_j}Y_j(g-ij,v) \notag\\
		&=Y^{w-MV}_i(g-ij,v)- \frac{w_i}{w_j} Y^{w-MV}_j(g-ij,v) \notag\\
		&=Y^{w-MV}_i(g,v)- \frac{w_i}{w_j} Y^{w-MV}_j(g,v).
	\end{align}

	Using {Component Balance}, we have $\sum_{i\in N(h)} \big[Y_i(g,v) - Y^{w-MV}_i(g,v) \big]=0$. \\
	Thus Eq.(\ref{eq5}) implies $\displaystyle\sum_{i\in N(h)} \frac{w_i}{w_j} \big[Y_j(g,v)-Y^{w-MV}_j(g,v)\big]=0$. \\
	This further implies that $\displaystyle\sum_{i \in N(h)}\frac{w_i}{w_j} n(h) \big [Y_j(g, v) - Y_j^{w-MV}(g,v)\big] = 0$. \\
	As $\dfrac{w_i}{w_j} > 0$ and $n(h) > 0$, so we have $Y_j(g,v)-Y^{w-MV}_j(g,v)=0$. \\
	It follows that $Y_j(g,v)=Y^{w-MV}_j(g,v)$. Thus, there exist an unique allocation rule satisfying {Component Balance} and {Weighted Bargaining Power}. 
	\end{proof}
\end{Theorem}
This characterization is an adaptation of the properties defined in \cite{Haeringer1999} whose interpretations are natural and can be extended directly to Network games. Our next axiom is \textit{Efficiency.}
\item \textit{Efficiency:}\label{axiom2}
An allocation rule $Y$ is \textit{Efficient} if for each $g \in \mathbb G^N$ and $v \in \mathbb V^N$ it holds that $$\sum_{i\in N(g)} Y_i(g, v) = v(g).$$
An {Efficient} network allocation rule distributes the total value generated by the network among the participating players.
\begin{lemma}\label{Lemma3}
	The weighted Myerson value satisfies {Efficiency}, namely, for all $g \in \mathbb G^N$, $w\in \mathbb W^N$, and $v \in \mathbb V^N$, we have
	\begin{equation*}
		\sum_{i\in N(g)} Y^{w-MV}_i(g, v) = v(g).
	\end{equation*}
\end{lemma}
	\begin{proof}
		Let $v\in \mathbb{V^{N}}$, $g\in \mathbb{G^{N}}$, and $w \in \mathbb W^N$. Then, we have
		\begin{align*}
		\sum_{i\in N(g)} Y_i^{w-MV}(g,v) &=\sum_{i\in N(g)}\sum_{g'\subseteq g:i\in N(g')} \dfrac{w_i}{\sum\limits_{j\in N(g^{'})} w_j}\lambda_{g'}(v)\\
		&=\sum_{g'\subseteq g:i\in N(g')} \dfrac{\sum_{i\in N}w_i}{\sum\limits_{j\in N(g^{'})} w_j}\lambda_{g'}(v)\\
	    &= \sum_{ g' \subseteq g}{\lambda_{g'}(v)}\\
		&=v(g).
		\end{align*}
		This completes the proof.
	\end{proof}	
\noindent Next we show that the weighted Myerson value satisfies {Additivity.}
	\begin{lemma}\label{lemma4}
		The weighted Myerson value satisfies {Additivity}.
\end{lemma}
\noindent {Additivity} follows directly from the fact that the unanimity coefficients $\lambda_{g'}(\cdot)$ in Eq.(\ref{Eq.5}) also satisfy {Additivity}, namely, $\lambda_{g'}(\alpha v + \beta u) = \alpha \lambda_{g'}(v) + \beta \lambda_{g'}(u)$ for $u, v \in \mathbb V^N$ and $\alpha, \beta \in \mathbb R$.\\
The next Lemma shows that the weighted Myerson value satisfies {Superfluous Link Property} due to \cite{belau3}. 
\begin{lemma}\label{lemma5}
The weighted Myerson value satisfies {Superfluous Link Property}.
\end{lemma}
\begin{proof}
For any $g\in \mathbb{G}^N$, $v\in \mathbb{V}^N$ and $w\in \mathbb{W}^N,$ let $l$ be a {Superfluous Link} in $g$ with respect to $v$. Then, we have, $v(g')=v(g'\setminus l)~~\forall g'\subseteq g$. Since $\forall S\subseteq N$, $g_{|_{S}}\subseteq g$, we have $$v(g_{|_{S}})=v(g_{|_{S}}\setminus l)=v((g\setminus l)_{|_{S}})~~\forall S\subseteq N.$$
Hence,
\begin{eqnarray*}
Y_i^{w-MV}(g,v)&=&\sum_{g'\subseteq g:i\in N(g')} \dfrac{w_i}{\sum\limits_{j\in N(g^{'})} w_j}\lambda_{g'}(v)\\
&=&\sum_{g'\subseteq g\setminus l:i\in N(g')} \dfrac{w_i}{\sum\limits_{j\in N(g^{'})} w_j}\lambda_{g'}(v)\\
&=&Y_i^{w-MV}(g\setminus l,v).
\end{eqnarray*} 
\end{proof}
In the following axiom, denoted as \textit{Network-Specific Player Anonymity}, we deviate from the conventional {Point Anonymity Property} of the Myerson value in Theorem \ref{them:Myerson2}. It is important to highlight that the value function is formulated such that players possibly generate a non-zero value only when the whole network is formed.
\begin{definition}\label{definition1}
A value function $v \in \mathbb{V}^N$ is called \textit{Network Specific Player Anonymous} in $g$ if $v(g_{|_{S}})= 0$ for all $S\subset N$.
\end{definition}
Thus, in a \textit{Network Specific Player Anonymous} game, the interplay among the players in smaller/sub networks is absent and therefore, their contributions are significant only when all the players in the given network contribute together. An ideal allocation rule for such a game should distinguish the allocation to two distinct players only by means of the weights they are endowed with. Hence, we have the following axiom. 
\item \textit{Network Specific Player Anonymity:}\\	
 An allocation rule $Y$ on  $\mathbb G^N  \times \mathbb V^N$ satisfies \textit{Network Specific Player Anonymity} if for every $g\in \mathbb G^N$, $v\in \mathbb V^N$ that is \textit{Network Specific Player Anonymous} on $g$, and $w \in \mathbb W^N$, there exists an $\alpha \in \mathbb{R}$ such that $$Y_i(g,v)=\alpha w_i  \;\textrm{for all $i \in N$.}$$
This axiom guarantees that players under the given situation receive payoffs proportional to their respective weights within the network.
	\begin{lemma}\label{lemma6}
	The weighted Myerson value satisfies {Network Specific Player Anonymity}.
	\end{lemma}
	\begin{proof} For any $g \in \mathbb G^N$ and $w\in \mathbb{W}^N$, let $v \in \mathbb V^N$  be {Network Specific Player Anonymous} in $g$. Then, $\lambda_{g'}(v) = 0$ for all $g'\subsetneq g$. It follows that there exists an $\beta \in \mathbb R$ such that 
	$\lambda_g(v) = v(g) = \beta$.
	Hence, we obtain,
	\begin{align*}
	Y_i^{w-MV}(g,v) &= \sum_{g' \subseteq g:i\in N(g')} \dfrac{w_i}{\sum\limits_{j\in N(g^{'})} w_j}\lambda_{g'}(v)\\
	&=\sum_{g' \subsetneq g:i\in N(g')} \dfrac{w_i}{\sum\limits_{j\in N(g^{'})} w_j}\lambda_{g'}(v)+ \dfrac{w_i}{\sum\limits_{j\in N(g)} w_j}\lambda_{g}(v)\\
	&=\dfrac{w_i}{\sum\limits_{j\in N(g)}w_j}\cdot \beta,\;\;\textrm{for each $i\in N(g)$}
	\end{align*}
	Now, by taking $\alpha = \dfrac{\beta}{\sum\limits_{j\in N(g)}w_j}$, we get the required expression.
\end{proof}
The following proposition will be useful for our next characterization of the weighted Myerson value.
\begin{proposition}\label{Proposition1}
If an allocation rule $Y$ on $G^N\times V^N$ is {Network Specific Player Anonymous} and {Efficient}, we have,
	{\[ Y_i(g,v)=\begin{cases} 
						\dfrac{w_i}{\sum_{j\in N(g)}w_j}v(g)& ~ \text{if} ~ i \in N(g)\\
						0& ~ \text{otherwise}.  
					\end{cases}
					\]}
\end{proposition}
\begin{proof}
Let, $v \in \mathbb V^N$ be {Network Specific Player Anonymous} and $g \in \mathbb G^N$ be given.  
\\For all $i \in N\setminus N(g)$, we have $Y_i(g,v)=0.$
For all $i\in N(g),$ we have,
$Y_i(g,v)=\alpha w_i$.
\\Using {Efficiency},
\begin{align*}
v(g)=\sum_{i\in N(g)}Y_i^{w-MV}(g,v)=\sum_{i\in N(g)}\alpha w_i
\end{align*}
So, we have, $$ \dfrac{v(g)}{\sum_{j\in N(g)} w_j}=\alpha\sum_{i\in N(g)}\dfrac{w_i}{\sum_{j\in N(g)} w_j}$$
and hence, we get, $$ \alpha=\dfrac{v(g)}{\sum_{j\in N(g)} w_j}.$$
Therefore, $$Y_i^{w-MV}(g,v)=\dfrac{w_i}{\sum_{j\in N(g)}w_j}v(g).$$
\end{proof}
Our second characterization follows.
\begin{Theorem}\label{theorem3}
The weighted Myerson value is the unique {Efficient} allocation rule on $\mathbb{G}^N\times \mathbb{V}^N$ that satisfies the {Network Specific Player Anonymity}, {Superfluous Player Property} and {Additivity.}
\end{Theorem}
\begin{proof}
From Lemma \ref{Lemma3}, \ref{lemma4}, \ref{lemma5} and \ref{lemma6}, it is evident that the weighted Myerson value satisfy {Efficiency},{Additivity}, {Superfluous Link Property} and {Network Specific Player Anonymity.} 
We now show the converse part. Our proof follows a similar procedure as that in Theorem 2 of \cite{belau3}. However, to ensure the thoroughness of the deductions, we describe it here in details again.\\
Let $Y^{w-MV}$ satisfy {Efficiency}, {Additivity}, {Superfluous Link Property} and {Network Specific Player Anonymity.} 
Let, $g\in  \mathbb G^N$, $v\in \mathbb{V}^{N}$, $w \in \mathbb W^N$. Recall from Eq.(\ref{eq1}), $v=\sum\limits_{\substack{g'\in \mathbb G^N\\ g'\neq g_0}} \lambda_{g'}(v)u_{g'}$ where $\lambda_{g'}(v)$ are unique unanimity coefficients  corresponding to unanimity value function $u_{g'}$.  By {Additivity} we have,	
	$$ Y^{w-MV}(g,v)=\sum\limits_{g'\subseteq g} Y^w(g, \lambda_{g'}(v)u_{g'}).$$
	
	Therefore, it is sufficient to show that $Y^{w-MV}(g,\lambda_{g'}(v)u_{g'})$ is uniquely determined for all $g'\in \mathbb G^N$. For $g'\in \mathbb G^N$, let $v_{g'}=\lambda_{g'}(v)u_{g'}.$
\\Case I: $g' \nsubseteq g$
\newline $v_{g'}(g'')=0 ~~\forall g''\subseteq g$. so, we have by Proposition \ref{Proposition1}, 

{\[ Y_i^{w-MV}(g,v_{g'})=\begin{cases} 
						0& ~ \text{if} ~ i \in N(g)\\
						0& ~ \text{if} ~ i \in N\setminus N(g).  
					\end{cases}
					\]}
Case II: $g'=g$
\newline $v_{g'}(g'')=0~~ \forall g''\subset g$ and $v_{g'}(g)=\lambda_{g'}(v)$, Proposition \ref{Proposition1} gives,
{\[ Y_i^{w-MV}(g,v_{g'})=\begin{cases} 
						\dfrac{w_i}{\sum_{j\in N(g)}w_j}\lambda_{g'}(v)& ~ \text{if} ~ i \in N(g)\\
						0& ~ \text{if} ~ i \in N\setminus N(g) .  
					\end{cases}
					\]}
Case III : $g' \subset g$
\newline Let, $l \in g\setminus g'$ and $g''\subseteq g$ such that $l\in g''$, then 
\begin{align*}
			v_{g'}(g'')&=v_{g'}(g''\backslash l)\\
					&=\begin{cases}
						\lambda_{g'}(v), \textrm{ if } g'\subseteq g''\\
						~~~~0, ~\textrm{otherwise}.
					\end{cases}
\end{align*}
\\Therefore, every link $l \in g\setminus g'$ is superfluous in $(g,v_{g'}).$  Thus, by the {Superfluous Link Property}  $\forall i \in N(g)$, we have
$Y^{w-MV}_i(g,v_{g'})= Y_i^{w-MV}(g\setminus l, v_{g'})$. Now, by continuous application of {Superfluous Link Property}, we have
$Y_i^{w-MV}(g,v_{g'})=Y_i^{w-MV}(g', v_{g'})$. As $(g',v_{g'})$ is {Network Specific Player Anonymous}, we have, 
{\[ Y_i^{w-MV}(g,v_{g'})= Y_i^{w-MV}(g',v_{g'})=\begin{cases} 
						\dfrac{w_i}{\sum_{j\in N(g)}w_j}\lambda_{g'}(v)& ~ \text{if} ~ i \in N(g')\\
						0& ~ \text{if} ~ i \in N\setminus N(g') .  
					\end{cases}
					\]}
\end{proof}
\begin{remark}
 The logical independence of each of the sets of axioms in Theorem \ref{theorem3} are shown below:
 \begin{itemize}
 \item $Y(g,v_o)=1$ for all $g\in \mathbb{G}^N$ and $v_o\in \mathbb{V}^N$ such that $v_o(g)=0~~\forall g\in \mathbb{G}^N$, satisfies {Additivity}, {Superfluous Link Property} and {Network Specific Player Anonymity} but not {Efficiency}.
 \item $Y(g,v)=\dfrac{v(g)}{n(g)}$ for all $g\in \mathbb{G}^N$ and $v\in \mathbb{V}^N$ satisfies {Efficiency}, {Additivity}, and {Network Specific Player Anonymity} but not {Superfluous Link Property}.
 \item $Y(g,v)=\alpha_iv(g)$ for all $g\in \mathbb G^N$ and $v\in \mathbb V^N$ such that $\alpha_i\neq \alpha_j$ for $i\neq j$, $i,j \in N$ and $\sum_{i\in N}\alpha_i=1$ and $\alpha_i\geq 0$ satisfies {Efficiency}, {Additivity}, and {Superfluous Link Property} but not {Network Specific Player Anonymity}.
 \item $Y(g,v)=Y^{MV}(g,v)$ if $v(g)\leq k$ and $Y(g,v)=Y^{PV}(g,v)\footnote{$Y^{PV}(g,v)$ is the position value for Network games discussed in \cite{Nouweland}.}$ if $v(g)>k$ for any $k \in \mathbb{R}$ satisfies {Efficiency}, {Network Specific Player Anonymity}, and {Superfluous Link Property} but not {Additivity}. 
 \end{itemize}
\end{remark}
\section{The w-potential function}
 The concept of the potential function for TU games is first proposed by Hart and Mas-Colell \cite{Hart}. Potential functions are mathematical tools used to analyze and characterize the behavior of TU games, particularly in terms of the allocation of the payoffs to the players. This concept finds further extensions in the works of Winter \cite{Winter}, Owen \cite{Owen}, Xu et al. \cite{Xu} etc., just to name a few. 
\newline Slikker \cite{Slikker05b} uses the notion of network potential to characterize the Myerson value. The network potential focuses on the player's overall marginal contributions through their links in the network. Thus, in the context of networks, the use of potential functions pertaining to TU games provides a powerful framework for understanding and analyzing various aspects of the interactions. 
\begin{definition}
Let, Let $g \in \mathbb G^N$ and $v \in \mathbb V^N$. Given a function $P:\mathbb G^N\times \mathbb V^N \rightarrow \mathbb{R}$ such that $P(g_o,v)=0$, the total marginal contribution of a player $i$ through all her links prescribed by $P$ is
$$D^{i}P(g,v)=P(g,v)-P(g\setminus g_i,v)~~~~~\forall (g,v)\in \mathbb G^N\times \mathbb V^N, i\in N(g). $$
\end{definition}
\begin{definition}
A function $P$ is said to be a player potential function if $P(g_o,v)=0,~~ \forall v \in \mathbb V^N$ and 
$$\sum_{i\in N(g)}D^{i}P(g,v)=v(g)~~~~~\forall (g,v)\in \mathbb G^N\times \mathbb V^N .$$
\end{definition}
\par In line with Slikker \cite{Slikker05b}, now we define a weighted version of the potential function which we call the weighted player potential function and define as follows.
\begin{definition}
Given a weight vector $w \in \mathbb W^N$, a function $P^{w}: \mathbb G^N \times\mathbb V^N \mapsto \mathbb R$ is said to be a weighted player potential function if $P^w(g_o,v)=0,~~ \forall v \in \mathbb V^N$ and 
$$\sum_{i\in N(g)}w_iD^{i}P^w(g,v)=v(g)~~~~~\forall (g,v)\in \mathbb G^N\times \mathbb V^N .$$
\end{definition}
The weighted player potential function is designed to assess the weighted contribution of a player in the network. It evaluates a player's influence in a network by considering their assigned weights.
\newline  We have the following theorem that characterizes the weighted Myerson value using player potential function.
\begin{Theorem}
For every positive weight vector $w \in \mathbb{W}^N$, there exists a unique weighted potential function $P^w$ in $\mathbb G^N\times \mathbb V^N$ such that $w_iD^{i}P^w(g,v)=Y_i^{w-MV}(g,v).$
\end{Theorem} 
\begin{proof}
Initially, we show the existence of a player potential function. Subsequently, we describe how the marginal contributions given by this player potential function is aligned with the weighted Myerson value.
\newline Let, $g\in \mathbb G^N$, $v \in \mathbb V^N$ and $w \in \mathbb W^N$ be given. Now, from Eq.(\ref{eq1}), we can write $v=\sum\limits_{\substack{g'\in \mathbb G^N\\ g'\neq g_0}} \lambda_{g'}(v)u_{g'}$.  We define $P^w: \mathbb G^N \times \mathbb V^N \mapsto \mathbb R$ as
\begin{equation}\label{Eq6}
P^w(g,v)=\sum_{g'\subseteq g}\dfrac{\lambda_{g'}(v)}{\sum\limits_{j\in N(g^{'})} w_j}~~~\forall g\in\mathbb{G}^N, v \in \mathbb{V}^N.
\end{equation}
Clearly, $P^w(g_o,v)=0.$ Also, $\forall g\in \mathbb{G}^N,$
\begin{align*}
D^{i}P^w(g,v)&=P^w(g,v)-P^w(g\setminus g_i,v)\\
&=\sum_{g'\subseteq g}\dfrac{\lambda_{g'}(v)}{\sum\limits_{j\in N(g^{'})} w_j}-\sum_{g'\subseteq g\setminus g_i}\dfrac{\lambda_{g'}(v)}{\sum\limits_{j\in N(g^{'})} w_j}\\
&=\sum_{g'\subseteq g:i\in N(g')}\dfrac{\lambda_{g'}(v)}{\sum\limits_{j\in N(g^{'})} w_j}\\
\implies w_iD^{i}P^w(g,v)&=w_i\sum_{g'\subseteq g:i\in N(g')}\dfrac{\lambda_{g'}(v)}{\sum\limits_{j\in N(g^{'})} w_j}\\
\implies w_iD^{i}P^w(g,v)&=\sum_{g'\subseteq g:i\in N(g')}w_i\dfrac{\lambda_{g'}(v)}{\sum\limits_{j\in N(g^{'})} w_j}\\
\implies w_iD^{i}P^w(g,v)&=Y_i^{w-MV}(g,v).
\end{align*}
Since $Y^{w-MV}$ satisfies {Efficiency}, it follows that  $\sum_{i \in N(g)}w_i D^i P^w( g, v) = v(g)$. Thus, we have for all $g \in \mathbb G^N$ and $v \in \mathbb V^N$, $P^w$ given by Eq.(\ref{Eq6}) is a weighted player potential function. 
\newline It remains to show that the weighted player potential function is unique. If $Q^w(g,v)$ is any weighted player potential function then we have that for all $(g,v)$ with $g\neq g_o$,
\begin{align}\label{Eq7}
v(g)&=\sum_{i\in N}w_i\Big[Q^w(g,v)-Q^w(g\setminus g_i,v)\Big]\nonumber\\
&=\sum_{i\in N(g)}w_i\Big[Q^w(g,v)-Q^w(g\setminus g_i,v)\Big]
\end{align}
The second equality holds since $Q^w(g,v)=Q^w(g\setminus g_i,v)$ for $g_i=g_0$ which is further equivalent to $i \in N_o(g).$
\newline Using Eq.(\ref{Eq6}). we have,
\begin{align}\label{Eq8}
&\sum_{i\in N(g)}w_i\Big[Q^w(g,v)-Q^w(g\setminus g_i,v)\Big]=v(g)\nonumber\\
&\implies \sum_{i\in N(g)}w_iQ^w(g,v)-\sum_{i\in N(g)}w_iQ^w(g\setminus g_i,v)=v(g)\nonumber\\
&\implies \sum_{i\in N(g)}w_iQ^w(g,v)=v(g)+\sum_{i\in N(g)}w_iQ^w(g\setminus g_i,v)\nonumber\\
&\implies Q^w(g,v)=\dfrac{1}{\sum_{i\in N(g)}w_i}\Big[v(g)+\sum_{i\in N(g)}w_iQ^w(g\setminus g_i,v)\Big].
\end{align}
For any $v \in \mathbb V^N$, we know that $Q^w(g_o,v)=0$. As a result, $Q^w(g,v)$ recursively using Eq.(\ref{Eq7}) proves the uniqueness of the weighted player potential function. This completes the proof.
\end{proof}
\end{itemize}
In the following section, we present the implementation of the weighted Myerson value.
\section{A bidding mechanism}\label{sec:5}
In this section, we study the non-cooperative framework of the weighted Myerson value, building upon the groundwork laid by \cite{Hart1996} and \cite{slikker07}. This non-cooperative foundation offers a strategic perspective to define the value, wherein cooperative value payoffs emerge from players' equilibrium strategies via a bidding mechanism. We describe two distinct mechanisms that implements the weighted Myerson value.
\subsection{Mechanism I}
 Our initial mechanism is an adaptation to Slikker's \cite{slikker07} implementation of the Myerson value under the conditions of sub-game perfect Nash equilibrium. Throughout this mechanism, we make the assumption that both the network $g \in \mathbb G^N$ and the underlying game $v \in \mathbb H^N$ are known to us. Our bidding process takes into consideration the weights of the players hence the only modification required is in the calculation of the weighted net bids. Other than this adjustment, the weighted bidding mechanism mirrors the steps outlined in \cite{slikker07}. To maintain completeness, we will outline the revised mechanism below and state the theorems without providing the proofs. 
\begin{definition}~\cite{slikker07}
A value $v \in \mathbb H^N$ is zero monotonic if  $v(g')-v(L_i(g'))\geq v(g'-L_i(g'))$ for $g' \subseteq g$ and $i \in N(g')$.
\end{definition} 
\subsubsection{Recursive formula}
Here, we introduce a recursive expression for the weighted Myerson value. This recursive formulation is deduced from the network potential function.
\begin{Theorem}\label{theorem4}
Let, $v\in \mathbb{V}^N$ and $g\in \mathbb G^N$. Then for all $i \in N(g)$, we have, 
\begin{equation}
Y^{w-MV}_i(g,v)=\frac{1}{\sum_{j \in N(g)}w_j}\left[ w_i[v(g)-v(g\backslash g_i)] +\sum\limits_{j \neq i} w_jY^{w-MV}_i(g\backslash g_j,v) \right].\footnote{If there is no potential for ambiguity, we will use the simplified notation $j\neq i$ instead of explicitly stating $j\in N(g)\setminus i$ throughout the rest of this paper.}
\end{equation}
\end{Theorem}
\begin{proof} It has already been shown in the previous section that 
\begin{align*}
Y_i^{w-MV}(g,v)
	&=w_iD^{i}P^w(g,v) \\
	&=w_i\Big[ P^w(g,v)-P^w(g\setminus g_i,v)\big] \\
	&=w_i \dfrac{1}{\sum_{j \in N(g)}w_j}\Big[v(g)+\sum_{j\in N(g)}w_jP^w(g\setminus g_j,v)-\sum_{j\in N(g)}w_jP^w(g\setminus g_i,v)\Big]\\
	&=\dfrac{1}{\sum_{j \in N(g)}w_j}\Big[w_iv(g)+\sum_{j\neq i}w_j\Big\{w_iP^w(g\setminus g_j,v)-w_iP^w(g\setminus \{g_i,g_j\},v)-w_iP^w(g\setminus g_i,v)\\
	&+w_iP^w(g\setminus \{g_i,g_j\},v)\Big\}\Big]\\
	&= \dfrac{1}{\sum_{j \in N(g)}w_j}\Big[w_iv(g)+\sum_{j\neq i}w_jY^{w-MV}_i(g\setminus g_j,v)-w_i\sum_{j\neq i}Y^{w-MV}_j(g\setminus g_i,v)\Big]\\
	&=\dfrac{1}{\sum_{j \in N(g)}w_j}\Big[w_i[v(g)-v(g\setminus g_i)]+\sum_{j\neq i}w_jY^{w-MV}_i(g\setminus g_j,v)\Big]\\
	\end{align*}
	\end{proof}
In the following sub-section, we will summarize the procedure for adapting our bidding mechanism to achieve the weighted Myerson value as an equilibrium outcome. 
\subsubsection{The bidding process}
 When $|g|=0$, it implies that the network has no links, resulting in the sole player $i$ receiving his individual standalone value of 0.
Now, let's consider the scenario where $|g|=k \geq 1$, and assume that the mechanism has been previously specified for networks and components containing a maximum of $k-1$ links. The mechanism operates through a series of rounds, and each round consists of three distinct stages (labeled as $t=1-3$). After the completion of a round, the game either concludes or initiates new rounds for any remaining players.
Now, we will thoroughly explain one round of the mechanism, specifically looking at a random network $g$ with $k$ links.

 \begin{enumerate}
 
  \item[\textbf{Stage 1.}] In this round, each player $i \in N(g)$ makes a bid {$b^{i}_j \in \mathbb{R}$} to all $j \in N(g) \setminus i$. \\
Let {$$B^{i}=\sum_{j\neq i} w_ib_{j}^{i}-\sum_{j\neq i}w_jb_{i}^{j}$$} represent the net bid of player $i$, which quantifies their 'relative' willingness to take on the role of the proposer.
\newline Let $i^{*}$ denote the player with the highest net bid in this round. In the event of multiple players having the same maximum value, any of these top bidders is selected as the 'winner' with equal probability.
\newline Moving on to the subsequent stage, the player $i^{*}$ selected as the "winner" in this manner becomes the proposer. \textbf{Proceed to Stage 2}.
\item[\textbf{Stage 2.}] The player $i^{*}$, who has been identified as the "winner" in the previous stage, now puts forward proposed payoffs denoted as $y_j$ to the other players $j \in N(g) \setminus {i^{*}}$. \textbf{Proceed to stage 3}.
\item[\textbf{Stage 3.}] In this stage, players apart from $i^{*}$ individually consider and sequentially decide whether to accept or reject the proposed offers. Consequently, two distinct cases emerge.
\begin{enumerate}
\item[\textbf{Case (a)}.] If the proposed offer is declined by at least one of the players, the players belonging to the set $N(g)$ move on to engage in the subsequent round. In this next round, the set of links within component $g$ becomes $|g\setminus g_{i^{*}}|.$ It is noteworthy that this ensuing round continues till the count of links in $|g\setminus g_{i^{*}}|\geq 1.$~~~~~~\textbf{Stop}.

\item[\textbf{Case (b)}.] If the proposed offer is accepted, then each player $j \in N(g) \setminus {i^{*}}$ receives a payoff of $y_{j}$, while player $i^{*}$ obtains $$v(g)-\sum_{j \in N(g) \setminus i^{*}} y_{j}.$$
\\ It is worth noting that these payoffs are received in addition to the bids made during stage 3, and possibly including payoffs from prior rounds. Consequently, the ultimate payoff for player $j \in N(g) \setminus {i^{*}}$ can be expressed as $$y_{j}+b_{j}^{i^{*},l}.$$
Similarly, the final payoff for player $i^{*}$ is given by $$v(g)-\sum_{j \in N(g) \setminus i^{*}}(y_{j}+b_{j}^{i^{*},l}).\;\;\;\;\;\textbf{ Stop}.$$ 
\end{enumerate}
\end{enumerate}
The mechanism described above applied to situation $(g,v)$ will be denoted by $\Gamma^{w-MV}(g,v).$
\newline Note that, the removal of players at the end of a round, also leads to the complete deletion of all links associated with the players which may result in the emergence of several components and, subsequently, separate parallel mechanisms. 
\newline Next we present the Theorems without their proofs.
\begin{Theorem}\label{theorem7}
For any $g\in \mathbb G^N$ and a zero-monotonic Network game $v \in \mathbb H^N$, there exists a sub-game perfect Nash equilibrium within the bidding mechanism where the players' payoffs align with their weighted Myerson value.
\end{Theorem}
In a sub-game perfect Nash equilibrium, the theorem shows that a player's payoff always matches her weighted Myerson value. This remains true even when the proposer is selected randomly. So, in this equilibrium, the outcome is guaranteed to be the player's weighted Myerson value.
\begin{Theorem}\label{theorem8}
For all $g\in \mathbb G^N$, a zero monotonic Network game $v \in \mathbb H^N$ and each sub-game perfect Nash equilibrium in $\Gamma^{w-MV}(g,v)$, the payoffs to the players coincides with the weighted Myerson value.
\end{Theorem}
In the following section, we present another implementation of the weighted Myerson value.
\subsection{Mechanism II}
In our implementation, we follow the bargaining process introduced by Hart and Mas-Colell in \cite{Hart1996}. We demonstrate that the expected outcomes in the bargaining equilibria correspond to the weighted Myerson values. This conclusion takes into consideration the possibility that the game may conclude without any agreements, factoring in a probability element.
\newline It is important to note that the fundamental concept of removing players at the end of a round aligns with the idea of deleting all the links associated with a player in this mechanism as well. However, it is essential to highlight that while their findings have some relevance, applying their results directly is not straightforward. This is primarily because removing a player's links could potentially divide the network into several components as the process continues. Consequently, this situation might result in the existence of several parallel mechanisms in subsequent rounds, which is distinct from the mechanism introduced by Hart and Mas-Colell \cite{Hart1996}, where such a scenario does not arise.
\subsubsection{Recursive formula}
We now introduce a recursive expression for the weighted Myerson value. 
\begin{Theorem}\label{theorem9}
Let $v\in \mathbb{V}^N$ and $g\in \mathbb G^N$. Then for all $i \in N$ and $S\subseteq N$, we have, 
\begin{equation}\label{Eq9}
Y^{w-MV}_i(g_{|_{S}},v)=\frac{1}{\sum_{j \in S}w_j}\left[ w_i[v(g_{|_{S}})-v(g_{|_{S}}\backslash g_i)] +\sum\limits_{j \neq i} w_jY^{w-MV}_i(g_{|_{S}}\backslash g_j,v) \right].
\end{equation}
\end{Theorem}
\begin{proof} Now,
\begin{align*}
&\frac{1}{\sum_{j \in S}w_j}\left[ w_i[v(g_{|_{S}})-v(g_{|_{S}}\backslash g_i)] +\sum\limits_{j \neq i} w_jY^{w-MV}_i(g_{|_{S}}\backslash g_j,v) \right]\\
&=\frac{w_i}{\sum_{j \in S}w_j} [v(g_{|_{S}})-v(g_{|_{S}}\backslash g_i)] +\frac{1}{\sum_{j \in S}w_j}\sum\limits_{j \neq i} w_jY^{w-MV}_i(g_{|_{S}}\backslash g_j,v)\\
&=\frac{w_i}{\sum_{j \in S}w_j}\sum_{g'\subseteq g_{|_{S}}: i\in N(g')}\lambda_{g'}(v)+\frac{1}{\sum_{j \in S}w_j}\sum_{j\neq i}w_j\sum_{g'\subseteq g_{|_{S}}\setminus g_j: i\in N(g')}\frac{w_i}{\sum_{j \in N(g')}w_j}\lambda_{g'}(v)\\
&=\sum_{g'\subseteq g_{|_{S}}: i\in N(g')}\frac{w_i}{\sum_{j \in N(g')}w_j}\lambda_{g'}(v)\\
&=Y^{w-MV}_i(g_{|_{S}},v).
\end{align*}
\end{proof}
Next, we proceed with the formal explanation of the mechanism. 
\subsubsection{The bidding process}
When the number of players in the network $g$ is 1,i.e., $n(g)=1$, the solitary player $i$ within this network receives her standalone value 0.
\par Now, let us consider the scenario where $n(g) = m$, and the mechanism is defined for networks or components consisting of at most $m-1$ players. The mechanism works in a series of rounds, and each round has three steps (denoted as $t = 1-3$). After a round is done, the game can either end or start new rounds for any remaining players.
\newline Now that we have set the context, we explain how one round works, designed for a network with $m$ players. It is also important to mention that there is a constant parameter, denoted as $0 \leq \rho < 1$, which is predetermined for the mechanism.
\newline The mechanism progresses in multiple rounds, and each round has different phases. In each round, a group of active participants, represented as $S \subseteq N$, is selected. From this group, one individual, denoted as $i \in S$, becomes the proposer. The various steps of the bargaining process within these rounds is as follows:
\begin{enumerate}
\item[\textbf{Stage 1.}]  In the initial round, the active set encompasses all players, i.e., $S = N$. The proposer is selected in a random manner from within the set $S$, with every player in $S$ having an equal likelihood of being chosen. Suppose, $i \in S$ is chosen as the proposer. \textbf{Proceed to Stage 2}. 
\item[\textbf{Stage 2.}] The proposer subsequently presents an offer, $$\sum_{j \in N}a^{g_{|_{S}},i}_j\leq v(g_{|_{S}}).$$ \textbf{Proceed to Stage 3}.
\item[\textbf{Stage 3.}] During this phase, participants within the set $S$, excluding the proposer $i$, individually assess and sequentially determine their acceptance or rejection of the proposed offers. As a result, two separate scenarios can occur.
\begin{enumerate}
\item[\textbf{Case (a)}.] If every member belonging to the set $S$ agrees to the proposed offer, ascertained through a predetermined sequence of inquiries, the game concludes, and the associated payoffs are determined accordingly.~~~~~~\textbf{Stop}.

\item[\textbf{Case (b)}.] Should the offer face rejection from a single member within the set $S$, a situation unfolds where, with a likelihood of $\rho$, progression proceeds to the subsequent round, encompassing the same active player set $S$. Conversely, with a probability of $1-\rho$, a breakdown scenario ensues: the proposer $i$ exits the game, obtaining a payoff of zero, and the group of active players transforms into $S\setminus i$.~~~~~~\textbf{Stop}. 
\end{enumerate}
\end{enumerate}
This process constitutes a sequential game with perfect information, wherein each step follows a predetermined order, and all players possess complete knowledge of previous actions. Within this context, a stationary sub-game perfect equilibrium (referred to as SP equilibrium) is established. Notably, the weighted average of proposals within this equilibrium aligns with the weighted Myerson value. Additionally, as the probability denoted by $\rho$ approaches unity, the proposals within these equilibria tend towards the weighted Myerson value, further reinforcing their convergence.
\par In this mechanism, it is important to highlight that the decisions made in a specific phase during a particular round can take into account the payoffs from previous rounds and stages, and this will not affect the overall analysis. To keep things clear in our notation, we will not always explicitly mention the exact specification that we are considering. Instead, we will concentrate on comparing payoffs within a specific setup.
\newline Additionally, it is important to consider that in the event of a breakdown, with a probability of $1-\rho$, when the player exits the network, it can potentially fragment into multiple distinct components. Consequently, multiple parallel mechanisms may also initiate as a result of this fragmentation.
\begin{proposition}\label{proposition2}
Let, $g\in \mathbb G^N$ and $v\in \mathbb H^N$ be zero monotonic. Then for each specification of $\rho(0\leq\rho<1)$, there is a SP equilibrium. The proposals corresponding to a SP equilibrium are always accepted and they are characterized by the following:
\begin{enumerate}
\item[\textbf{a.1}] $a^{g_{|_{S}},i}_i(\rho)=v(g_{|_{S}})-\sum\limits_{j\neq i}a^{g_{|_{S}},i}_j(\rho)$ for each $i\in S\subseteq N$.
\item[\textbf{a.2}]$a^{g_{|_{S}},i}_j(\rho)=\rho a^{g_{|_{S}}}_j(\rho)+(1-\rho)a^{g_{|_{S\setminus i}}}_j(\rho)$ for each $i,j\in S$ with $i\neq j$ and $S\subseteq N$ where $$a^{g_{|_{S}}}(\rho)=\dfrac{1}{\sum\limits_{i\in S}w_i}\sum\limits_{i\in S}w_ia^{g_{|_{S}},i}(\rho).$$ Moreover, these proposals are unique and non-negative.
\end{enumerate}
\end{proposition}
We use the recursive formula given in the next proposition and omit the formal proof of this proposition as it easily follows along the lines of the proof of Proposition 2 in \cite{Hart1996}. 
\begin{proposition}\label{proposition3}
Let, $(a^{g_{|_{S}}}(\rho))_{S\subseteq N}$ be the payoff configuration associated with
proposals satisfying (a.1) and (a.2), then it holds that
\begin{equation}\label{Eq10}
a^{g_{|_{S}}}_i(\rho)=\frac{1}{\sum_{j \in S}w_j}\left[ w_i[v(g_{|_{S}})-v(g_{|_{S\setminus i}})] +\sum\limits_{j \neq i} w_ja^{g_{|_{S\setminus j}}}_i(\rho)\right], i\in S\subseteq N.
\end{equation}
Moreover, these vectors $(a^{g_{|_{S}}}(\rho))_{S\subseteq N}$ are unique and non-negative.
\end{proposition}
\begin{proof}
Let, $g\in \mathbb G^N$ and $v\in \mathbb H^N$ be zero monotonic. By (a.1), for any $i\in S\subseteq N$, we have,
\begin{align*}
\sum_{j \in S}w_ja^{g_{|_{S}}}_i(\rho)&=w_i\Big(v(g_{|_{S}})-\sum_{j\neq i}a^{g_{|_{S}},i}_j(\rho)\Big)+\sum_{j\neq i}w_ja^{g_{|_{S}},j}_i(\rho)
\end{align*}
Applying (a.2),
\begin{align}
\sum_{j \in S}w_ja^{g_{|_{S}}}_i(\rho)&=w_i\Big(v(g_{|_{S}})-\sum_{j\neq i}\Big\{\rho a^{g_{|_{S}}}_j(\rho)+(1-\rho)a^{g_{|_{S\setminus i}}}_j(\rho)\Big\}\Big)\nonumber \\
&~~~~~~~~~~~~~~~~~~~~~~~+\sum_{j\neq i}w_j\Big\{\rho a^{g_{|_{S}}}_i(\rho)+(1-\rho)a^{g_{|_{S\setminus j}}}_i(\rho)\Big\}\nonumber\\
&=w_iv(g_{|_{S}})-\sum_{j\neq i}\Big\{w_i\rho a^{g_{|_{S}}}_j(\rho)+(1-\rho)w_ia^{g_{|_{S\setminus i}}}_j(\rho)\Big\} \nonumber\\ &~~~~~~~~~~~~~~~~~~~~~~~+\sum_{j\neq i}\Big\{\rho w_ja^{g_{|_{S}}}_i(\rho)+(1-\rho)w_ja^{g_{|_{S\setminus j}}}_i(\rho)\Big\}\nonumber\\
&=w_iv(g_{|_{S}})-\rho\sum_{j\neq i}w_ia^{g_{|_{S}}}_j(\rho)-(1-\rho)\sum_{j\neq i}w_ia^{g_{|_{S\setminus i}}}_j(\rho)\nonumber\\
&~~~~~~~~~~~~~~~~~~~~~~~+\rho\sum_{j\neq i}w_j a^{g_{|_{S}}}_i(\rho)+(1-\rho)\sum_{j\neq i}w_ja^{g_{|_{S\setminus j}}}_i(\rho)\nonumber\\
&=w_iv(g_{|_{S}})-w_iv(g_{|_{S\setminus i}})+\sum_{j\neq i}w_ja^{g_{|_{S\setminus j}}}_i(\rho)\nonumber
\end{align}
Thus, we have $$a^{g_{|_{S}}}_i(\rho)=\dfrac{1}{\sum_{j \in S}w_j}\Big[w_i\Big(v(g_{|_{S}})-v(g_{|_{S\setminus i}})\Big)+\sum_{j\neq i}w_ja^{g_{|_{S\setminus j}}}_i(\rho)\Big].$$
The payoff of the single coalitions $\{i\}$ is $a^{g_{|_{\{i\}}}}_i(\rho)=v(\{i\})=0$ (i.e., $i\in N_0(g))=Y^{w-MV}_{i}(\{i\},v)$ for all $i\in N.$ 
Therefore, the non-negativity and the uniqueness of $a^{g_{|_{S}}}_i(\rho)$ follows from the monotonicity of $v$ and Eq.(\ref{Eq10}) applied recursively.
Hence, from Eq.(\ref{Eq9}), we conclude that $a^{g_{|_{S}}}_i(\rho)=Y^{w-MV}_i(g_{|_{S}},v)~~\forall S\subseteq N.$
\end{proof}
\begin{Theorem}\label{theorem10}
Let, $g\in\mathbb G^N$ and $v\in\mathbb H^N$ be zero monotonic. Then for each $0\leq\rho<1$, there is a unique SP equilibrium. Moreover, for all $i\in S\subseteq N$, the SP equilibrium payoff vector $(a^{g_{|_{S}},i}(\rho))_{S\subseteq N}$ coincides with the weighted Myerson value for $(g_{|_{S}},v).$ Also, as $\rho\rightarrow 1$, the SP equilibrium proposals $(a^{g_{|_{S}}}_i(\rho))_{S\subseteq N, i\in S}$ converges to the weighted Myerson value for $(g_{|_{S}},v).$
\end{Theorem}
\begin{proof}
\textcolor{blue}{The first part of the theorem follows directly from Propositions \ref{proposition2} and \ref{proposition3}. Now, for $\rho\rightarrow 1$, from (a.1) and (a.2), we have,
\begin{align*}
a^{g_{|_{S}},i}_i(\rho)&=v(g_{|_{S}})-\sum_{j\neq i}a^{g_{|_{S}},i}_j(\rho)\\
&=v(g_{|_{S}})-\sum_{j\neq i}[\rho a^{g_{|_{S}}}_j(\rho)+(1-\rho)a^{g_{|_{S\setminus i}}}_j(\rho)]\\
&=(1-\rho)v(g_{|_{S}})-(1-\rho)v(g_{|_{S\setminus i}})+\rho a^{g_{|_{S}}}_i(\rho)\\
&=(1-\rho)[v(g_{|_{S}})-v(g_{|_{S\setminus i}})]+\rho a^{g_{|_{S}}}_i(\rho)\\
\end{align*}
and 
\begin{align*}
a^{g_{|_{S}},j}_i(\rho)=\rho a^{g_{|_{S}}}_i(\rho)+(1-\rho)a^{g_{|_{S\setminus j}}}_i(\rho).
\end{align*}
which implies that $a^{g_{|_{S}},i}_i(\rho)\rightarrow a^{g_{|_{S}}}_i(\rho)$ and $a^{g_{|_{S}},j}_i(\rho)\rightarrow a^{g_{|_{S}}}_i(\rho).$}
\end{proof}
\section{Conclusion}
In this paper, we begin by presenting characterizations of the weighted Myerson value within Network games. This formulation closely aligns with the weighted Myerson value previously defined by Haeringer \cite{Haeringer1999} for communication structures. Furthermore, we introduce the notion of w-potential in a network structure. Another novelty is the bidding mechanism of weighted Myerson values, employing a procedure similar to the one outlined earlier by Slikker \cite{slikker07} and Hart and Mas-Colell \cite{Hart1996} in networks and cooperative TU-games respectively. Through our analysis, we demonstrate that the weighted Myerson value finds support within each sub-game perfect Nash equilibrium of the individualistic game outlined in section 5.
\par By introducing the bidding mechanism, we gain valuable insights into the interactions that takes place among players. This approach provides a framework for comprehending the determinants that underlie the existence of this specific set of allocation rules. It is worth noting that this allocation rule can potentially be extended to games situated on multigraphs as well as hypergraphs, a direction we intend to explore in our forthcoming research endeavors.

\end{document}